# Quantum-Mechanical Correlations and Tsirelson Bound from Geometric Algebra

Carsten Held[1]

(Dated: November 10, 2020)

The Bell-Clauser-Horne-Shimony-Holt inequality can be used to show that no local hidden-variable theory can reproduce the correlations predicted by quantum mechanics (QM). It can be proved that certain QM correlations lead to a violation of the classical bound established by the inequality, while all correlations, QM and classical, respect a QM bound (the Tsirelson bound). Here, we show that these well-known results depend crucially on the assumption that the values of physical magnitudes are scalars. The result implies, first, that the origin of the Tsirelson bound is geometrical, not physical; and, second, that a local hidden-variable theory does not contradict QM if the values of physical magnitudes are vectors.

Keywords: Bell-CHSH inequality, Tsirelson bound, non-locality, geometric algebra

The Bell-Clauser-Horne-Shimony-Holt (Bell-CHSH) inequality [1] is a well-known generalization of the Bell inequality [2] and, like its famous predecessor, is designed to prove that no hidden-variable theory can reproduce the correlations predicted by quantum mechanics (QM) if it respects an assumption of locality [3, 4]. It can be shown that some QM correlations lead to a violation of the classical bound established by the inequality and it can also be shown that no correlations, QM or classical, exceed a specific QM bound, i.e. the Tsirelson bound. Here, we show that these well-known results depend crucially on the assumption that the values of physical magnitudes are scalars. More specifically, the assumption that these values are not scalars, but vectors that are elements of the geometric algebra $\mathbf{G}^3$ over $\mathbf{R}^3$, makes it possible that the classical bound is violated and the QM bound respected, even given a locality assumption. The result implies, first, that the origin of the Tsirelson bound is geometrical, not physical; and, second, that a local hidden-variable theory does not contradict QM if the values of physical magnitudes are vectors in the geometric algebra $\mathbf{G}^3$.

The Bell-CHSH inequality and the argument based on it are well-known (see e.g. [5]), so a brief introduction of both will suffice here. Consider a large collection of pairs of spin-½ particles measured for their spin components in certain directions of $\mathbf{R}^3$. Consider four unit vectors $\mathbf{a}, \mathbf{a}', \mathbf{b}, \mathbf{b}' \in \mathbf{R}^3$, such that $\mathbf{a}$ and $\mathbf{a}'$ ($\mathbf{b}$ and $\mathbf{b}'$) define two measurement directions for the left (right) particle. Let the oriented axis defined e.g. by vector $\mathbf{a}$ be identical with the physical magnitude A (which in QM becomes the observable A) and let the magnitude's value found upon measurement be either 1 or – 1, and analogously for $\mathbf{a}', \mathbf{b}, \mathbf{b}'$. Consequently, the expectation value E ($\mathbf{ab}$) of the product of values of A and B, revealed in a measurement of spin in the $\mathbf{a}$-direction on the left and the $\mathbf{b}$-direction on the right, is an element of [– 1, 1], and similarly for other combinations of the four vectors. We define the *classical correlation* c ($\mathbf{a}, \mathbf{b}$) as such an expectation value: c ($\mathbf{a}, \mathbf{b}$) = E ($\mathbf{ab}$) = $\int_\Lambda$ (E($\mathbf{a}\,\mathbf{b}$)|λ) dP (λ), where λ is a variable describing an eventual joint influence on the values of A and B. To derive the desired inequality, we make three assumptions. First, we assume *locality*, i.e. we assume that an expectation value of the form E ($\mathbf{ab}$|λ) factorizes: E ($\mathbf{ab}$|λ) = E ($\mathbf{a}$|λ) E ($\mathbf{b}$|λ). Second, we assume that |E ($\mathbf{x}$|λ)| ≤ 1, for $\mathbf{x} = \mathbf{a}, \mathbf{a}', \mathbf{b}$ or $\mathbf{b}'$, something we will refer to as *boundedness*. Third, we note the simple *algebraic fact* that |x + y| + |x – y| ≤ 2 for real numbers x, y ∈ [– 1, 1]. Consider now the four classical correlations c($\mathbf{a}, \mathbf{b}$), c($\mathbf{a}, \mathbf{b}'$), c($\mathbf{a}', \mathbf{b}$), and c($\mathbf{a}', \mathbf{b}'$). Using the above definition, we have:

|c($\mathbf{a}, \mathbf{b}$) + c($\mathbf{a}, \mathbf{b}'$) + c($\mathbf{a}', \mathbf{b}$) – c($\mathbf{a}', \mathbf{b}'$)| ≤

$\int_\Lambda$ |E($\mathbf{ab}$|λ) + E($\mathbf{ab}'$|λ)| + |E($\mathbf{a}'\mathbf{b}$)|λ) − (E($\mathbf{a}'\mathbf{b}'$|λ)| dP (λ).  (1)

We abbreviate the RHS of (1) as $\int_\Lambda$ **S** (λ) dP (λ) and consider the integrand **S** (λ) separately, suppressing the λ-dependence. Then, using our three assumptions in turn, we get:

$$\begin{aligned}
\mathbf{S} &= |E(\mathbf{ab}) + E(\mathbf{ab}')| + |E(\mathbf{a}'\mathbf{b}) - E(\mathbf{a}'\mathbf{b}')| \\
&= |E(\mathbf{a})|(|E(\mathbf{b}) + E(\mathbf{b}')|) + |E(\mathbf{a}')|(|E(\mathbf{b}) - E(\mathbf{b}')|) \quad \text{(locality)} \\
&\leq |E(\mathbf{b}) + E(\mathbf{b}')| + |E(\mathbf{b}) - E(\mathbf{b}')| \quad \text{(boundedness)} \\
&\leq 2. \quad \text{(algebraic fact)}
\end{aligned}$$
(2)

---

[1] E-mail: carsten.held@mailbox.org

From (1) and (2), we now have the Bell-CHSH inequality:

$$|c(\mathbf{a}, \mathbf{b}) + c(\mathbf{a}, \mathbf{b}') + c(\mathbf{a}', \mathbf{b}) - c(\mathbf{a}', \mathbf{b}')| \leq 2. \tag{3}$$

Inequality (3) is violated by suitable QM systems, as is easily shown. Define an observable for the spin component in the direction of vector **a** as a self-adjoint operator $A = \sigma^1 \mathbf{a}$ on a two-dimensional Hilbert space $\mathbf{H}^A$ (where $\sigma^1$ is the spin operator on $\mathbf{H}^A$ acting on the state vector representing system $A_n$), with $\|A\| \leq 1$ (where '$\|\ldots\|$' is the operator norm). Similarly, for vector **a'** define operator $A' = \sigma^1 \mathbf{a}'$ on $\mathbf{H}^A$, and analogously, for vectors **b**, **b'**, define operators $B = \sigma^2 \mathbf{b}$ and $B' = \sigma^2 \mathbf{b}'$ on another two-dimensional space $\mathbf{H}^B$ for system $B_n$. Define the *QM correlation* $\langle A, B\rangle := \langle \sigma^1 \mathbf{a} \otimes \sigma^2 \mathbf{b}\rangle_\psi$ for QM state $\psi \in \mathbf{H}^A \otimes \mathbf{H}^B$. Assuming that $\psi$ is the singlet state, we calculate $\langle A, B\rangle = -\cos\theta_{ab}$, and analogously for $\langle A, B'\rangle$, $\langle A', B\rangle$, $\langle A', B'\rangle$.

The classical and QM correlations do not match. Assume, for reductio, that correlations of both kinds lead to inequalities exhibiting the same bound, i.e. (3) becomes:

$$|\langle A, B\rangle + \langle A, B'\rangle + \langle A', B\rangle - \langle A', B'\rangle| \leq 2. \tag{4}$$

Assume, second, that **a**, **a'**, **b**, **b'** are coplanar with angles:

$$\sphericalangle(\mathbf{a}, \mathbf{a}') = \sphericalangle(\mathbf{b}', \mathbf{b}) = \pi/2 \text{ and } \sphericalangle(\mathbf{a}', \mathbf{b}') = \pi/4. \tag{5}$$

Then, the QM correlations are calculated to be: $\langle A, B\rangle = \langle A, B'\rangle = \langle A', B\rangle = -\langle A', B'\rangle = +1/\sqrt{2}$, such that the LHS of (4) equals $4/\sqrt{2} = 2\sqrt{2}$. Thus, given (5), (4) implies $2\sqrt{2} \leq 2$, a contradiction. Instead of the naively assumed (4), QM yields, again given angles defined as in (5):

$$|\langle A, B\rangle + \langle A, B'\rangle + \langle A', B\rangle - \langle A', B'\rangle| = 2\sqrt{2}. \tag{6}$$

Accordingly, QM correlations generally do not obey the Bell-CHSH inequality. (Henceforth, the argument leading to the contradiction of (4) and (6) is referred to as the Bell-CHSH argument; and the situation containing many pairs of spin-½ particles in the singlet state as the Bell-CHSH situation.)

However, the violation of (4) by QM cannot be arbitrarily large. It was shown by Tsirelson [6] that, for any choice of unit vectors **a**, **a'**, **b**, **b'**:

$$|\langle A, B\rangle + \langle A, B'\rangle + \langle A', B\rangle - \langle A', B'\rangle| \leq 2\sqrt{2}. \tag{7}$$

The RHS of (7) is the quantum-mechanical (or Tsirelson) bound for the violation of the Bell-CHSH inequality. We sketch a simple proof of (7) for the special case $\|A\| = \|A'\| = \|B\| = \|B'\| = 1$. Define an operator $\mathbf{B} = AB + AB' + A'B - A'B'$. (7) is proved if $|\langle\mathbf{B}\rangle| \leq 2\sqrt{2}$. We have:

$$\mathbf{B}^2 = 4 \cdot \mathbf{1} - \mathbf{C}, \tag{8}$$

where **1** is the unit operator and $\mathbf{C} = [A, A'][B, B']$. (Note that the derivation of (8) requires both $[A, A'] \neq 0 \neq [B, B']$ and $[A, B] = [A, B'] = [A', B] = [A', B'] = 0$.) Consider now the real number $\|\mathbf{C}\|$. Since $\|[A, A']\| \leq 2$, and analogously for $B$, $B'$, we immediately have:

$$\|\mathbf{C}\| \leq 4. \tag{9}$$

From (8) and (9), we get:

$$\|\mathbf{B}\| = \sqrt{(\|4 \cdot \mathbf{1} - \mathbf{C}\|)} \leq \sqrt{(4 + \|\mathbf{C}\|)} \leq \sqrt{(4 + 4)} = 2\sqrt{2} \tag{10}$$

and finally, from (10) ($\|\mathbf{B}\| \leq 2\sqrt{2}$), we get $|\langle\mathbf{B}\rangle| \leq 2\sqrt{2}$ such that (7) is proved.

The derivations of (6) and (7) make essential use of two basic concepts of QM: the definition of QM correlations from Hilbert space operators and the non-commutation of some of these operators (i.e.: $[A, A'] \neq 0 \neq [B, B']$). However, there is a way to obtain these very results without making any quantum-mechanical assumptions and instead employing elementary Geometric Algebra (GA). This alternative derivation is of theoretical interest as it uses GA elements (vectors in $\mathbf{R}^3$ and constructions from them) but no QM machinery (no operators acting on vectors in $\mathbf{H}^A \otimes \mathbf{H}^B$) and thus opens a way to a new, geometric understanding of (6) and (7).

Our GA approach to (6) and (7) proceeds in two steps. First, we interpret the unit vectors **a**, **a'**, **b**, **b'** $\in \mathbf{R}^3$ as elements of a geometric algebra. Second, we allow the *values* of physical magnitudes to be identical with these vectors. We begin with the introduction of a geometric algebra. Let $\{\mathbf{e_1}, \mathbf{e_2}, \mathbf{e_3}\}$ be an orthonormal basis of $\mathbf{R}^3$ generating $\mathbf{G}^3$, the geometric algebra over $\mathbf{R}^3$, by means of the geometric product [7, 8, 9]. The unit vectors **a**, **a'**, **b**, **b'** $\in \mathbf{R}^3$ now can be written in terms of the basis vectors as $\mathbf{a} = a_1\mathbf{e_1} + a_2\mathbf{e_2} + a_3\mathbf{e_3}$, etc. The basis vectors instantiate a characteristic non-commutative structure, given by $\mathbf{e_i}\mathbf{e_j} + \mathbf{e_j}\mathbf{e_i} = 2\delta_{ij}$ ($i, j = 1, 2, 3$), that carries over to other vectors: generally, neither one of the vectors **a**, **a'**, **b**, **b'** commutes with any other. Consider, on the other hand, that we are interested in a GA account of the Bell-CHSH situation, where the Pauli operators describing the spin in QM exhibit the same non-commutative structure as the vectors $\mathbf{e_1}$, $\mathbf{e_2}$, $\mathbf{e_3}$ in $\mathbf{G}^3$. We would thus expect this structure to be the critical factor in our GA approach – but it turns out that this is not the case. Although the structure arguably is crucial in GA accounts of other (non-statistical) no-hidden-variables arguments [10], it does not play any role in a GA description of the Bell-CHSH situation. Here, as we will see presently, the assumption of values of magnitudes as vectors is all-important.

We thus assume that the values of the physical magnitudes referred to in the Bell-CHSH argument (i.e. spin components) are vectors. In particular, we assume that the unit vectors **a**, **a'**, **b**, **b'** introduced above are possible values of such magnitudes and thus that functions of these values are functions of vectors.

We write the functions as 'F(...)', writing a function of vector **x** (or vectors **x**, **y**) as F(**x**) (or F(**x**, **y**)). Recall the classical expectations employed in (2) above. In a single value expectation, say E(**a**), the vector **a** represents a physical magnitude A identified with the oriented axis spanned by **a**. The values ± **a** of A are interpreted as values of a random variable taking dichotomic values and are identified with ± 1. Accordingly, E(**a**), E(**b**), etc. are scalars and E(**ab**), etc. are sums of products of scalars. In the new proposal, the argument **a** of F(**a**) also is a vector representing a physical magnitude A identified with the oriented axis spanned by **a**; the values ± **a** of A are also interpreted as values of a random variable taking dichotomic values. However, these values now keep their identification with ± **a** (instead of being replaced by ± 1). We assume that F(**a**), a function of vector **a**, is itself a vector on the axis spanned by **a**, or, more specifically, that F(**a**) = α **a**, where α ϵ [– 1, 1], and analogously for **a′**, **b**, **b′**. Now, consider a new scalar **S′**, structurally similar to **S** in (2), and define it as:

$$\mathbf{S'} = |F(\mathbf{a}, \mathbf{b}) + F(\mathbf{a}, \mathbf{b'})| + |F(\mathbf{a'}, \mathbf{b}) - F(\mathbf{a'}, \mathbf{b'})|. \quad (11)$$

Starting from (11), we repeat the first two steps of (2), i.e. we assume locality (factorizability) for F(**a**, **b**), etc. and boundedness for F(**a**), F(**a′**). Thus, we get:

$$\mathbf{S'} \leq |F(\mathbf{b}) + F(\mathbf{b'})| + |F(\mathbf{b}) - F(\mathbf{b'})|$$
$$= |\alpha \mathbf{b} + \beta \mathbf{b'}| + |\alpha \mathbf{b} - \beta \mathbf{b'}|, \quad (12)$$

where α, β ϵ [– 1, 1]. Suppose first that α = β = 1 such that F(**b**) = **b** and F(**b′**) = **b′** and thus:

$$\mathbf{S'} \leq |\mathbf{b} + \mathbf{b'}| + |\mathbf{b} - \mathbf{b'}| \quad (13)$$

Recall the elementary algebraic fact, used in (2), that, if b, b′ ϵ [– 1, 1], then |b + b′| + |b – b′| ≤ 2. Unsurprisingly, this fact does not carry over to our GA approach. It is not the case that generally |**b** + **b′**| + |**b** – **b′**| ≤ 2; instead, it is the case that |**b** + **b′**| + |**b** – **b′**| = 2 iff **b** = ± **b′**.

Due to the fact that **b** and **b′** are vectors, the term on the RHS of (13) can be larger than 2. Assume again that **a**, **a′**, **b**, **b′** form the angles, by now familiar, from (5). Due to our choice that **a** = **e₁** and **a′** = **e₂**, these angles imply that **b** = – (**e₁** + **e₂**)/√2, **b′** = (– **e₁** + **e₂**)/√2. Thus, **b** + **b′** = – √2 **e₁** and **b** – **b′** = – √2 **e₂**. Thus, we have |**b** + **b′**| = |**b** – **b′**| = √2 and (13) becomes:

$$\mathbf{S'} \leq |\mathbf{b} + \mathbf{b'}| + |\mathbf{b} - \mathbf{b'}| = 2\sqrt{2}. \quad (14)$$

Comparing (2) and (14), we see that **S** and **S′** both have been evaluated using a locality assumption – but (2) sets a stricter bound for **S** than (14) does for **S′**. In particular, the derivation in (2) does not extend to the case for which **S′** was introduced: a GA account of the Bell-CHSH situation, wherein values of the physical magnitudes involved are vectors, not scalars. Using (14) and (6) we cannot establish a contradiction comparable to the one between (4) and (6). In brief, **S′** is not in a logical conflict with QM predictions, although it is classical in the sense that it respects a locality assumption.

(14) has been derived for unit vectors **b**, **b′**, specified as in (5). It is clear, however, that (14) extends to arbitrary unit vectors. The term |**b** + **b′**| + |**b** – **b′**| is maximal if |**b** + **b′**| = |**b** – **b′**| = √2. Thus, for arbitrary unit vectors **b**, **b′**:

$$\mathbf{S'} \leq |\mathbf{b} + \mathbf{b'}| + |\mathbf{b} - \mathbf{b'}| \leq 2\sqrt{2}. \quad (15)$$

Finally, we consider the general case: α, β ϵ [– 1, 1] and arbitrary unit vectors **a**, **a′**, **b**, **b′**. We distinguish three special cases. First, assume αβ > 0. Then, | α **b** + β **b′** | ≤ |**b** + **b′**| and | α **b** – β **b′** | ≤ |**b** – **b′**|. Second, assume αβ < 0. Then, | α **b** + β **b′** | ≤ |**b** – **b′**| and | α **b** – β **b′** | ≤ |**b** + **b′**|. Thus, in both cases:

$$|\alpha \mathbf{b} + \beta \mathbf{b'}| + |\alpha \mathbf{b} - \beta \mathbf{b'}| \leq |\mathbf{b} + \mathbf{b'}| + |\mathbf{b} - \mathbf{b'}|. \quad (16)$$

From (15) and (16):

$$|\alpha \mathbf{b} + \beta \mathbf{b'}| + |\alpha \mathbf{b} - \beta \mathbf{b'}| \leq 2\sqrt{2}. \quad (17)$$

Finally, assume αβ = 0, e.g. α = 0; then the LHS of (16) equals 2 |β| ≤ 2. Thus, (17) follows for all three cases and we have the unconditional inequality:

$$\mathbf{S'} \leq 2\sqrt{2}. \quad (18)$$

All in all, we have shown that, for arbitrary α, β ϵ [– 1, 1] and unit vectors **a**, **a′**, **b**, **b′** ϵ $G^3$, the sum **S′** is bounded by 2√2. As we see from (14), this bound is attained when α = β = 1 and the unit vectors form angles as prescribed in (5) above. Since **S′** ≤ 2√2 – in contrast with **S** ≤ 2 – the Bell-CHSH inequality cannot be derived from it and thus the Bell-CHSH argument cannot be repeated for **S′**.

The F-functions have been introduced above to produce an expression **S′** that is the *classical* **S**, but with scalars replaced by vectors. These functions, it turns out, are closely related to the *quantum-mechanical* expectations. We have assumed that F(**a**) = α **a**, where α ϵ [– 1, 1], and analogously for **a′**, **b**, **b′**, which entails the boundedness of F(**a**) and F(**a′**). Moreover, we may assume that F(**a**, **b**) = α β **a b**, which, given the preceding assumptions, is equivalent to locality (factorizability). With these specifications, we can derive (12) from (11) without extra assumptions. Now we make two further assumptions, first that F(**a**, **b**) is a scalar and second that α = – β = 1. As a result, we get F(**a**, **b**) = – **a** · **b** = – cos θ_{ab}, which is just the QM correlation <A, B>. (Analogously for the other F-functions.) The QM correlations thus appear as special cases of the assumptions we made for the F-functions. It is thus no wonder that from these functions we can derive counterparts (14) and (18) of the QM results (6) and (7) – effectively: that we can derive these QM results not from QM, but from GA. The crucial background assumption of these derivations is that **a**, **a′**, **b**, **b′** in (11) are vectors, not scalars.

The significance of our result for the interpretation of QM is twofold. First, it is often assumed that the existence of an upper bound $2\sqrt{2}$ for the Bell-CHSH expression (the LHS of (7)), is a specific characteristic of QM that must be explained from *physical* principles. But the derivations in (11-18) imply something else, i.e. that this bound is of purely *geometric* origin, being due solely to the particular arrangement of the four $\mathbf{R}^3$ unit vectors **a**, **a′**, **b**, **b′**. To be sure, the crucial QM assumption made in the derivation of (7) is that $[A, A'] \neq 0 \neq [B, B']$, but this is not a QM-specific, or even specifically physical, assumption, as witnessed by the fact that its GA counterpart $[\mathbf{a}, \mathbf{a'}] \neq 0 \neq [\mathbf{b}, \mathbf{b'}]$ follows directly from (a) writing **a**, **a′**, **b**, **b′** in terms of the basis vectors $e_1, e_2, e_3$, (b) the assumptions $\mathbf{a} \neq \pm \mathbf{a'}$ and $\mathbf{b} \neq \pm \mathbf{b'}$, and (c) the non-commutativity of $e_1, e_2, e_3$. (It is claimed [11] that the geometric origin of the Tsirelson bound is proved already in [12], in a different context.)

The second, much weightier, implication concerns the Bell-CHSH argument for a disproof of local hidden variable theories for QM. A locality assumption is explicitly present in (2) and thus in (4), and it apparently is the only assumption in the whole argument that can reasonably be doubted. This has, in the past, led to the oft-repeated claim that the argument as a whole constitutes a general proof of the nonlocality of QM (see e.g. [13]). However, it is clear now that this claim requires careful qualification.

As we saw, the first two steps of the three-step derivation of (2) and the two steps summarized in (12) include the same locality (factorizability) assumption. This assumption is employed before the classical and GA derivations diverge. Hence, there is reason to infer that the GA account is just as local as the classical one. This, however, implies that the GA account respects the locality assumption and nevertheless matches the QM predictions summarized in (6) and (7). Hence, this account casts the unqualified opposition between locality and QM predictions into doubt. An explicit contradiction of both has been exhibited above, but it arises only given the tacit presupposition that the values of physical magnitudes are scalars; without that presupposition it can be avoided.

We have seen derivations of QM correlations (in (14)) and the Tsirelson bound (in (18)) using only the key idea of values of physical magnitudes as vectors. These derivations show that there is no general, entirely unconditional, conflict between QM correlation predictions and locality; the contradiction we have is conditional on the premise that the values of physical magnitudes are scalars, not vectors. Certainly, at some point we have to ask whether the idea of vectors as values is at all plausible. However, what is at issue here is not plausibility, but the co-existence of QM correlations and locality – which is enabled by the new idea. Moreover, at first glance the idea *is* plausible. The Bell-CHSH situation involves vectorial physical magnitudes (spin components) and in this context the assumption that these magnitudes have vectorial values, unfamiliar as it may be, makes good sense.

More questions suggest themselves, among them the following: Can the GA account of the Bell-CHSH situation be adapted to non-statistical no-hidden-variables arguments like the Greenberger-Horne-Zeilinger argument [14, 15]? This question requires a separate treatment and a recent proposal [10] has not yet answered it satisfyingly. Further research is required.

———————————————————————